# Reentrant Metallic Behavior in the Weyl Semimetal NbP

J. Xu[1,2], D. E. Bugaris[1], Z. L. Xiao*[1,2], Y. L. Wang[1,3], D. Y. Chung[1], M. G. Kanatzidis[1,4], and W. K. Kwok[1]

[1]Materials Science Division, Argonne National Laboratory, Argonne, Illinois 60439, USA
[2]Department of Physics, Northern Illinois University, DeKalb, Illinois 60115, USA
[3]Department of Physics, University of Notre Dame, Notre Dame, Indiana 46556, USA
[4]Department of Chemistry, Northwestern University, Evanston, Illinois 60208, USA

We report the occurrence of reentrant metallic behavior in the Weyl semimetal NbP. When the applied magnetic field $H$ is above a critical value $H_c$, a reentrance appears as a peak in the temperature dependent resistivity $\rho_{xx}(T)$ at $T = T_p$, similar to that observed in graphite where it was attributed to local superconductivity. The $T_p(H)$ relationship follows a power-law dependence $T_p \sim (H - H_c)^{1/\nu}$ where $\nu$ can be derived from the temperature dependence of the zero-field resistivity $\rho_0(T) \sim T^\nu$. From concurrent measurements of the transverse $\rho_{xx}(T)$ and Hall $\rho_{xy}(T)$ magnetoresistivities, we reveal a clear correlation between the rapidly increasing $\rho_{xy}(T)$ and the occurrence of a peak in the $\rho_{xx}(T)$ curve. Quantitative analysis indicates that the reentrant metallic behavior arises from the competition of the magnetoconductivity $\sigma_{xx}(T)$ with an additional component $\Delta\sigma_{xx}(T) = \kappa_H \sigma_{xx}(T)$ where $\kappa_H = [\rho_{xy}(T)/\rho_{xx}(T)]^2$ is the Hall factor. We find that the Hall factor ($\kappa_H \approx 0.4$) at peak temperature $T_p$ is nearly field-independent, leading to the observed $T_p(H)$ relationship. Furthermore, the reentrant metallic behavior in $\rho_{xx}(T)$ also is reflected in the behavior of $\rho_{xx}(H)$ that ranges from non-saturating at $T > 70$ K to saturation at liquid helium temperatures. The latter can be explained with the magnetic field dependence of the Hall factor $\kappa_H(H)$. Our studies demonstrate that a semiclassical theory can account for the 'anomalies' in the magnetotransport phenomena of NbP without invoking an exotic mechanism.

Weyl semimetals of the transition-metal monophosphides TaAs, TaP, NbAs and NbP [1-21] are new exotic topological quantum materials that have recently attracted extensive attention. They host Weyl fermions in the bulk and feature Fermi arcs on the surfaces [2]. Their bulk and surface band structures have been directly revealed through angle-resolved photoemission spectroscopy [2,3,5-7,10,12,14,15]. Experiments on electronic transport, which is crucial for potential applications in electronic devices, have uncovered novel properties such as ultra-high mobility [4,8], extremely large magnetoresistance (XMR) [4,8], and chiral-anomaly-induced negative magnetoresistance [9,11,13,21].

Analogous to topological insulators [22-25], the surface states of Weyl semimetals could exhibit unusual quantum phenomena such as weak anti-localization induced by quantum interference [26]. Furthermore, weak anti-localization and localization [26] could arise from Weyl fermions in the bulk [1,9,11,26,27], highlighting the likely prospects for these materials as platforms for exploring novel phenomena. Indeed, saturating magnetoresistances (MR), which typically do not occur in non-Weyl XMR materials, have recently been observed in NbAs [21] and NbP [4, 18] crystals at liquid helium temperatures. They could be an indication of contributions from the topological surfaces that could have a saturating MR due to weak anti-localization and/or from the disordered bulk, which could have both weak anti-localization and localization and could induce a negative MR [26]. The temperature dependence of the MR in TaAs crystals [11] shows a reentrant metallic behavior similar to that observed in graphite [28], which has been attributed to the magnetic-field induced occurrence of local superconductivity. Recently, tip induced superconductivity was reported in TaAs [29] and surface superconductivity in NbAs through selective ion sputtering [30].

Here, we report magnetotransport investigations on the Weyl semimetal NbP. We observed both reentrant metallic temperature behavior and saturating magnetoresistivity. We show that these transport phenomena can be understood using a semiclassical theory, without the need to invoke exotic mechanisms such as magnetic-field induced local superconductivity [28] or weak (anti-)localization arising from quantum interference [26,27]. We demonstrate that the Hall factor – a quantity that is conveniently determined from the measured Hall ($\rho_{xy}$) and transverse ($\rho_{xx}$) magnetoresisitivities - plays a critical role in the understanding of both the reentrant metallic behavior and the magnetoresistivity saturation in transition-metal monophosphide Weyl semimetals and other materials such as graphite. In particular, the Hall factor depicts the deviation of $\rho_{xx}$ from its potential maximum value enabled by the magnetoconductivity $\sigma_{xx}$. Furthermore, its temperature and magnetic field dependences are responsible for the reentrant metallic behavior and the magnetoresistivity saturation, respectively.

Two NbP crystals were grown following the procedure outlined in Ref.4. DC resistivity measurements in constant current mode were conducted using a Quantum Design PPMS (PPMS-9). The magnetic field $H$ was applied along the *c*-axis of the crystal with the current *I* flowing in the *a-b* plane. We measured $\rho_{xx}(H)$ and $\rho_{xy}(H)$ at various stable temperatures and constructed the resistivity versus temperature $\rho_{xx}(T)$ and $\rho_{xy}(T)$ curves. Both crystals showed similar results and here we present results from sample A.

As expected for a semimetal, the zero-field resistivity $\rho_0$ given in the inset of Fig.1(a) for NbP decreases when the temperature is lowered. However, in the presence of an external magnetic field *H*, the temperature dependence of the resistivity becomes complicated, as presented in Fig.1(a). At $H \leq 1$ T, the temperature dependence of the magnetoresistivity $\rho_{xx}(T)$ at $T < 100$ K can become semiconductor-like, i.e. a magnetic field-induced suppression of the metallic behavior occurs. At larger fields, the metallic temperature dependence re-appears, following the semiconductor-like behavior, leading to a clearly defined peak in all $\rho_{xx}(T)$ curves above $H$ = 1.5 T. The magnetic-field induced reentrant metallic behavior in $\rho_{xx}(T)$ is reminiscent of similar response first observed in graphite [28]. Likewise, as shown in Fig.2(a), the magnetic field dependence of the peak temperature $T_p(H)$ also follows a power-law relationship $T_p \sim (H - H_c)^{1/\nu}$, with $\nu \approx 3$, similar to that reported for graphite ($\nu \approx$ 2-4) [28,31,32].

Since its discovery in graphite in 2003, the reentrant metallic behavior has been observed both in non-topological semimetals including Bi [32] and Sb [33] as well as recently in topological semimetals $Bi_{0.96}Sb_{0.04}$ [34], TaAs [11,35] and $WP_2$ [36]. In the first report in graphite [28], this phenomenon was attributed to the occurrence of local superconductivity induced by quantum Hall effect in the quantum limit. However, this mechanism, cannot account for the reentrant metallic behavior observed in other materials mentioned above, whose quantum limits were not reached in the reported experiments. Recently, the claim on 'hidden superconductivity' in graphite was questioned by Forgan [37], although he could not give alternative explanation. Thus, the reentrant metallic behavior in topological and non-topological semimetals remains an unsolved puzzle.

In a semi-classical theory, for a magnetic field $H$ applied in the $z$-direction and current flowing along the $x$-axis, the magnetoresistivity tensor is given as [38]:

$$\hat{\rho} = \begin{pmatrix} \rho_{xx} & \rho_{yx} \\ \rho_{xy} & \rho_{yy} \end{pmatrix} \quad (1)$$

where $\rho_{xx} = \sigma_{yy}/[\sigma_{xx}\sigma_{yy} + (\sigma_{xy})^2]$, $\rho_{yy} = \sigma_{xx}/[\sigma_{xx}\sigma_{yy} + (\sigma_{xy})^2]$, $\rho_{yx} = -\rho_{xy} = \sigma_{xy}/[\sigma_{xx}\sigma_{yy} + (\sigma_{xy})^2]$. $\sigma_{ij}$ ($i, j = x, y$) are the components of the magnetoconductivity tensor:

$$\hat{\sigma} = \begin{pmatrix} \sigma_{xx} & \sigma_{yx} \\ \sigma_{xy} & \sigma_{yy} \end{pmatrix} \quad (2)$$

with $\sigma_{xx} = ne\mu_x/(1 + \mu_x\mu_y H^2)$; $\sigma_{yy} = ne\mu_y/(1 + \mu_x\mu_y H^2)$; $\sigma_{yx} = -\sigma_{xy} = ne\mu_x\mu_y H/(1 + \mu_x\mu_y H^2)$ for an electron pocket. Here $n$ is the electron density, and $\mu_x$ and $\mu_y$ are the respective mobilities along the $x$ and $y$ axes. For a hole pocket, $\sigma_{ij}$ can be obtained by changing the sign of both the charge $e$ and the mobility. It can also be implemented for the isotropic case by assuming $\mu_x = \mu_y$.

NbP has four Fermi pockets with two pairs for the electrons and holes, respectively [18]. To calculate its magnetoresistivity, $\sigma_{ij}$ in Eq.(2) needs to be replaced with $\sigma_{ij} = \sum_p \sigma_{ij}^p$ where $p$ represents all the Fermi pockets. Since $\sigma_{ij}$ for each Fermi pocket has three free variables ($n$, $\mu_x, \mu_y$), it is nearly impossible to conduct a quantitative analysis of the magnetoresistivity for NbP by directly using Eq.(1). Currently, the popular analysis approach is to adopt an isotropic two-band model [4,8,11], i.e., Eq.(1) is simplified by assuming that the density of electrons and holes are compensated, all Fermi pockets are spherical, and the electrons and holes from different Fermi pockets have the same mobility $\mu_e$ and $\mu_h$, respectively. As demonstrated recently for the MR in YSb [38] and LaSb [39], the outcome of such an analysis for a multiband material could be unreliable.

We can re-formulate Eq.(1) to avoid the need of detailed information on individual Fermi pockets by utilizing experimentally accessible parameters. From Eqs.(1) and (2) we have $\sigma_{xx} = \rho_{yy}/(\rho_{xx}\rho_{yy} + \rho_{xy}^2)$. The crystal symmetry in NbP leads to $\sigma_{xx} = \sigma_{yy}$, thus $\rho_{xx} = \rho_{yy}$ and $\sigma_{xx} = \rho_{xx}/(\rho_{xx}^2 + \rho_{xy}^2)$. Then, the relationship for $\rho_{xx}$ and $\sigma_{xx}$ can be written as:

$$\rho_{xx} = 1/(\sigma_{xx} + \kappa_H \sigma_{xx}) \qquad (3)$$

with $\kappa_H = (\rho_{xy}/\rho_{xx})^2$, which we term as the Hall factor. Eq.(3) shows that the measured magnetoresistivity $\rho_{xx}$ is determined by the magnetoconductivity $\sigma_{xx}$ together with an additional term $\Delta\sigma_{xx} = \kappa_H \sigma_{xx}$ contributed by the Hall effect. Since $\kappa_H \geq 0$, thus $\rho_{xx} \leq 1/\sigma_{xx}$.

Furthermore, $\sigma_{xy}$ and $\rho_{xy}$ have opposite signs for electrons and holes, as given in Eqs.(1) and (2). Thus, the values of $\sigma_{xy}$ and $\rho_{xy}$ could decrease with the addition of a Fermi pocket. In the extreme case where $\sigma_{xy}$ and $\rho_{xy}$ of electrons and holes compensate each other, $\kappa_H \approx 0$ and $\rho_{xx}$ reaches its maximum $\rho_{xx}^M = 1/\sigma_{xx}$. When $\kappa_H \neq 0$, Eq.(3) indicates that its temperature and magnetic field dependences will affect those of $\rho_{xx}$. That is, $\kappa_H$ can be the determining factor for the temperature behavior of $\rho_{xx}$.

Experimentally $\kappa_H$ can be conveniently obtained by simultaneously measuring $\rho_{xx}$ and $\rho_{xy}$. In Fig.1(b) we present the temperature dependence of $\kappa_H(T)$ obtained at various magnetic fields with the corresponding $\rho_{xy}$ given in the inset. Clearly, the reentrant metallic behavior in $\rho_{xx}$ is accompanied by a rapidly increasing $\kappa_H$ with decreasing temperature.

In Fig.3, we use the $\rho_{xx}(T)$ curve obtained at $H = 9$ T to elucidate the role of $\kappa_H$ on the reentrant metallic behavior. In the inset we present the calculated $\sigma_{xx}$, $\kappa_H \sigma_{xx}$ and their sum $\sigma_{xx} + \kappa_H \sigma_{xx}$. It shows that $\sigma_{xx}$ decreases monotonically as the temperature is lowered and its sum with the Hall effect induced magnetoconductivity, $\sigma_{xx} + \kappa_H \sigma_{xx}$, also follows the same behavior when the value of $\kappa_H$ is small at high temperatures. At lower temperatures ($T < 100$K), however, the temperature response of $\sigma_{xx}$ and $\kappa_H \sigma_{xx}$ diverges because of the rapid increase of $\kappa_H$ with decreasing temperature, resulting in a dip in the $\sigma_{xx} + \kappa_H \sigma_{xx}$ curve. This corresponds to the observed peak in $\rho_{xx}(T)$, which is the inverse of $\sigma_{xx} + \kappa_H \sigma_{xx}$. The upper limit ($\rho_{xx}^M$) of $\rho_{xx}(T)$ in the absence of the Hall contribution, i.e. $\kappa_H = 0$, is also presented in Fig.3 with a dashed green curve. It indicates that the reentrant metallic behavior occurs only when $\rho_{xx}(T)$ deviates significantly from $\rho_{xx}^M$. In fact, we can re-write Eq.(3) as $\rho_{xx} = \rho_{xx}^M/(1 + \kappa_H)$. That is, $\kappa_H$ is a direct measure of the deviation of $\rho_{xx}(T)$ from its upper limit. Its sufficiently large value as well as fast change with temperature give rise to this intriguing reentrant behavior.

In order to directly demonstrate the correlations between $\kappa_H(T)$ and the reentrant metallic behavior, we plot $\kappa_H(T)$ and normalized $\rho_{xx}(T)$ curves obtained at the same fields in the inset of Fig.2(b). It shows that the peak in $\rho_{xx}(T)$ occurs nearly at the same $\kappa_H$, independent of the magnetic fields. Data in the main panel of Fig.2(b) gives $\kappa_H \approx 0.4$ at $T_p$ for all given magnetic fields.

Without knowing the temperature dependence of the density and mobility of the charge carriers in each Fermi pocket in this multi-band material, one is unable to analytically derive the temperature dependence of the Hall factor $\kappa_H$ and its value at $T_p$. Our experimental finding of field-independence of $\kappa_H$ at $T_p$, however, could be useful in understanding the $T_p$ versus $H$ relationship, which was used as evidence for supporting the magnetic-field induced 'superconductivity' interpretation of the reentrant metallic behavior in graphite [28]. In the case of graphite, which can be described with an anisotropic two-band model, we obtain $\kappa_H = [(n_e - n_h)\mu_y H/(n_e + n_h)]^2$ using Eqs.(1)-(3) by assuming $\mu_x^e = \mu_x^h = \mu_x$ and $\mu_y^e = \mu_y^h = \mu_y$, where $e$ and $h$ in the subscript and superscript stand for electron and hole respectively. If $n_e$ and $n_h$ are independent of temperature or have similar temperature dependence, a constant $\kappa_H(T_p)$ implies that $\mu_y H$ is the same at different $T_p$, i.e., $\mu_y H = $ const. Since $\mu_y$ should behave the same as $\mu_x$, the temperature dependence of the mobility $\mu_y$ can be inferred from that of the zero-field resistivity $\rho_0(T) = 1/[(n_e + n_h)e\mu_x]$. This leads to $\mu_y(T) \sim 1/\rho_0(T)$. Thus, we have $H \sim 1/\mu_y(T_p) \sim \rho_0(T_p)$. For a Fermi liquid with $\rho_0(T) = a + bT^2$ we can deduce a power-law relationship for $T_p \sim (H - H_c)^{1/2}$ with $H_c$ related to $a$. This is exactly what was observed by Kopelevich et al. in graphite and bismuth [32]. The situation in a multi-band system should be more complicated. In our NbP, however, $T_p(H)$ also seems to correlate strongly with $\rho_0(T)$. Data in the inset of Fig.1(a) for our NbP crystal at $T < 100$ K indicates $\rho_0(T) = a + bT^\nu$ with $\nu = 3$, which is the same exponent derived from the power-law relationship of $T_p(H)$, as discussed above and shown in Fig.2(a). In general, $\rho_0(T)$ at low temperatures can be described as $\rho_0(T) = a + bT^\nu$ with $\nu = 2$ for electron-electron scattering (in Fermi liquid), $\nu = 3$ and 4 for electron-phonon scattering in the pure and dirty limit, respectively [40,41]. In XMR materials, $n$ is often found to be 2 [42-44] while $\nu = 3$ [45, 46] and $\nu = 4$ [47] are also reported. Thus, we can conclude that a power-law dependence may be a general behavior of $T_p(H)$, with the exponent that can be estimated from $\rho_0(T) = a + bT^n$. In fact, $T_p \sim (H - H_c)^{1/\nu}$ with $\nu = 4$ was also reported for graphite [31], although it was attributed to spin-orbit interactions. Thus, no exotic mechanism such as magnetic-field induced local superconductivity is needed to account for the observed power-law behavior of $T_p(H)$.

Accompanying the reentrant metallic temperature behavior we also observed significant variation in $\rho_{xx}(H)$, ranging from non-saturating at high temperature ($T > 70$K) to saturation at liquid helium temperatures, as shown in Fig.4. In a Weyl semimetal, quantum interference could induce weak-antilocalization on the surface and both weak-antilocalization and localization in the bulk [26]. Combined with the orbital magnetoresisitivity, this could result in a complicated $\rho_{xx}(H)$ behavior, as demonstrated by the multi-variable fitting of the saturating magnetoresistivity in Bi$_{0.97}$Sb$_{0.03}$ [26]. As shown in the inset of Fig.5(a), however, the *MR* of our NbP crystal is larger than $10^5$ % even at 40 K, excluding the possibility of (anti-)localization behavior in the crystal. In Fig.5 we demonstrate that the variation in $\rho_{xx}(H)$ is due to the magnetic field dependence of the Hall factor $\kappa_H(H)$. Applying the same procedure used in the analysis of $\rho_{xx}(T)$ in Fig.3, we show the calculated $\sigma_{xx}(H)$, $\kappa_H\sigma_{xx}$ and their sum $\sigma_{xx} + \kappa_H\sigma_{xx}$ in Fig.5(b). It indicates that at low fields the Hall effect plays a negligible role and the total magnetoconductivity ($\sigma_{xx} + \kappa_H \sigma_{xx}$) is nearly equal to $\sigma_{xx}$, while at high fields it saturates to a value determined by the Hall effect ($\kappa_H \sigma_{xx}$). We also plot $\kappa_H(H)$ in Fig.5(b). It indicates that a noticeable deviation of $\sigma_{xx} + \kappa_H \sigma_{xx}$ from $\sigma_{xx}$ occurs at $\kappa_H$ larger than 0.4, the value determined at $T_p$ (see inset of Fig.2(b)). Comparison of $\rho_{xx}(H)$ with its upper limit $\rho_{xx}^M$ in Fig.5(a) shows that the measured magnetoresistivity is only a fraction of its potential maximum (~1/5 at $H$ = 9 T). The plot in the inset of Fig.5(a) also indicates that MR in the absence of the Hall contribution follows a power-law behavior with an exponent (= 1.65) close to but less than 2, consistent with that observed in non-Weyl materials showing non-saturating XMR [38,39,48].

In summary, we observed reentrant metallic temperature behavior accompanied with magnetoresistivity saturation in the Weyl semimetal NbP. We quantitatively described the phenomena with a semi-classical theory and revealed that they originate from the temperature and magnetic-field dependence of the Hall factor, respectively. This work highlights a general approach to explain 'unusual' magnetotransport behaviors based on the experimentally obtained Hall factor.

**ACKNOWLEDGEMENTS**

The crystal growth and magnetoresistivity measurements were supported by the U.S. DoE, Office of Science, Basic Energy Sciences, Materials Sciences and Engineering Division. Theoretical studies were carried out at Northern Illinois University under NSF Grant No. DMR-1407175. Electrical contacting was conducted at Argonne's Center for Nanoscale Materials (CNM), which was supported by DOE BES under Contract No. DE-AC02-06CH11357.

*Corresponding author, xiao@anl.gov or zxiao@niu.edu

**Figure captions**

**FIG.1.** (color online) (a) Temperature dependence of the magnetoresistivity $\rho_{xx}(T)$ at $H = 0.5$ T to 9 T with intervals of 0.5 T. The inset presents the temperature dependence of the zero field resistivity $\rho_0(T)$. Symbols are experimental data and dashed line is a fit of $\rho_0(T) = a + bT^3$. (b) the corresponding Hall factor $\kappa_H = [\rho_{xy}(T)/\rho_{xx}(T)]^2$, with $\rho_{xy}(T)$ to be the Hall magnetoresistivity given in the inset. The symbols in (a), (b) and the inset of (b) are the same.

**FIG.2.** (color online) (a) and (b) show the magnetic field dependence of the peak temperature ($T_p$) and the Hall factor $\kappa_H$ at $T_p$, respectively. In (a) the symbols are experimental data and the line represents $T \sim (H-H_c)^{1/3}$. Inset of (a) presents the $T_p(H)$ data as $T^3 \sim H$ to directly show the power-law dependence. Inset of (b) contains data of $\kappa_H(T)$ (symbols) and the associated normalized magnetoresistivity (lines) at $H = 2T, 3T, 4T, 5T, 7T$ and $9T$, demonstrating that the Hall factor is nearly the same at the peak temperature for $\rho_{xx}(T)$ curves obtained in different magnetic fields starts at nearly

**FIG.3.** (color online) Temperature dependence of the measured $\rho_{xx}$ at $H = 9$ T and the calculated magnetoresistivity ($1/\sigma_{xx}$) without the contribution ($\Delta\sigma_{xx} = \kappa_H\sigma_{xx}$) from the Hall field. The inset shows the temperature dependence of the magnetoconductivity $\sigma_{xx}$, the Hall-field induced component $\kappa_H\sigma_{xx}$, and their sum $\sigma_{xx}+\kappa_H\sigma_{xx}$, demonstrating the competition of $\sigma_{xx}$ and $\sigma_{xx}+\kappa_H\sigma_{xx}$ in their temperature dependence and the occurrence of a dip in their sum.

**FIG.4.** (color online) Magnetic field dependence of (a) the transverse magnetoresistivity $\rho_{xx}(H)$, (b) the Hall magnetoresistivity $\rho_{xy}(H)$, and (c) the Hall factor $\kappa_H = (\rho_{xy}/\rho_{xx})^2$ at temperatures $T = 3K$ and 10 K to 70 K at intervals of 5 K.

**FIG.5.** (color online) (a) Comparison of the measured $\rho_{xx}(H)$ (symbols) and its upper limit $1/\sigma_{xx}(H)$ at $T = 40$ K. Its inset shows the data as $MR = (\rho_{xx} - \rho_0)/\rho_0$. The dashed line

indicates a power-law relationship with an exponent of 1.65. (b) Magnetic field dependence of the calculated magnetoconductivity $\sigma_{xx}$, the additional component $\Delta\sigma_{xx} = \kappa_H \sigma_{xx}$ contributed by the Hall field and their sum. $\kappa_H(H)$ is also plotted in (b), which indicates that the deviation of $\rho_{xx}(H)$ to $1/\sigma_{xx}(H)$ becomes noticeable at $\kappa_H > 0.4$.

**FIGURE 1**

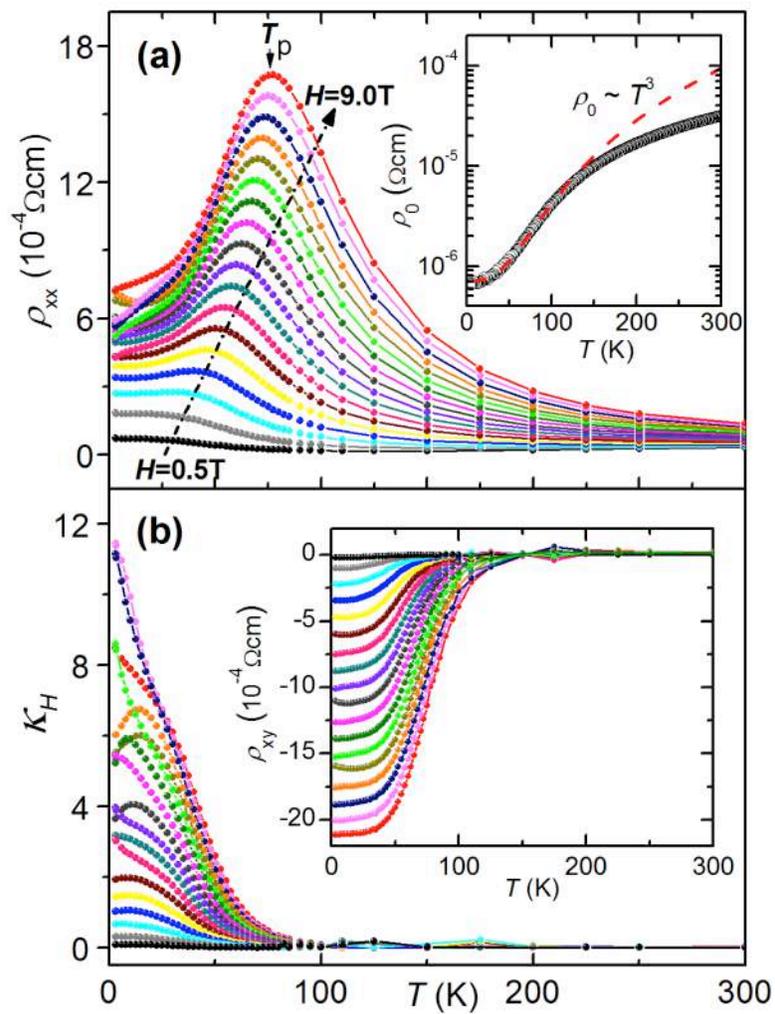

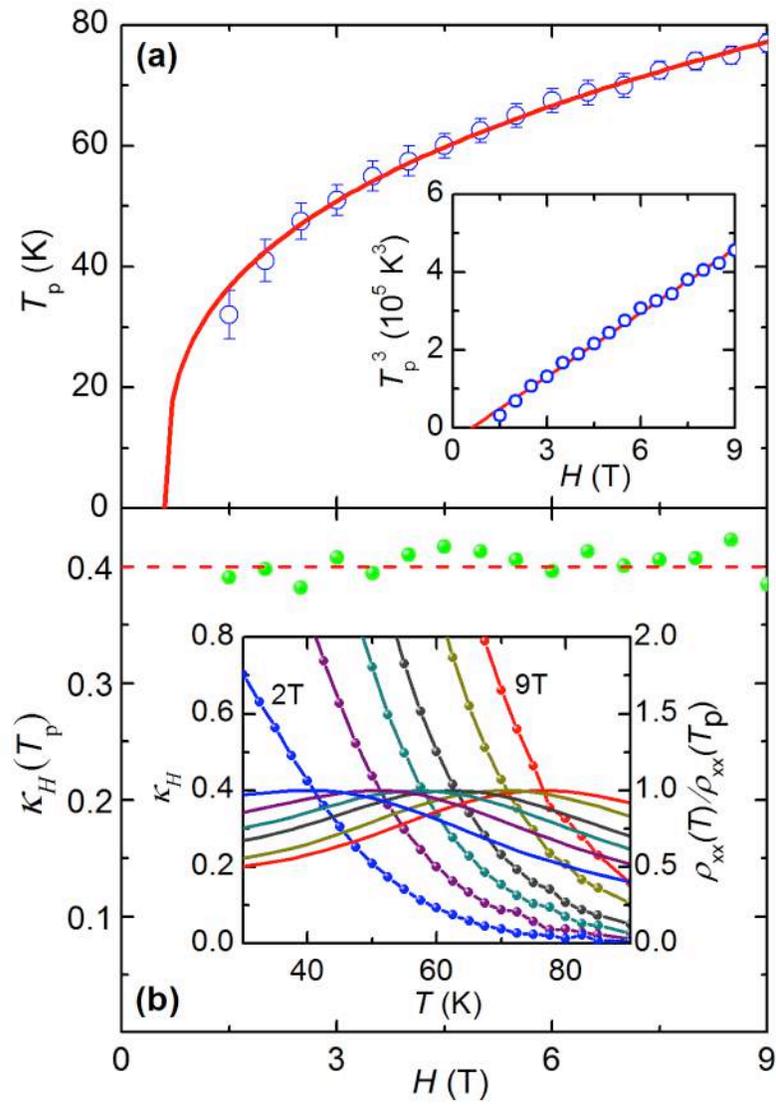

**FIGURE 3**

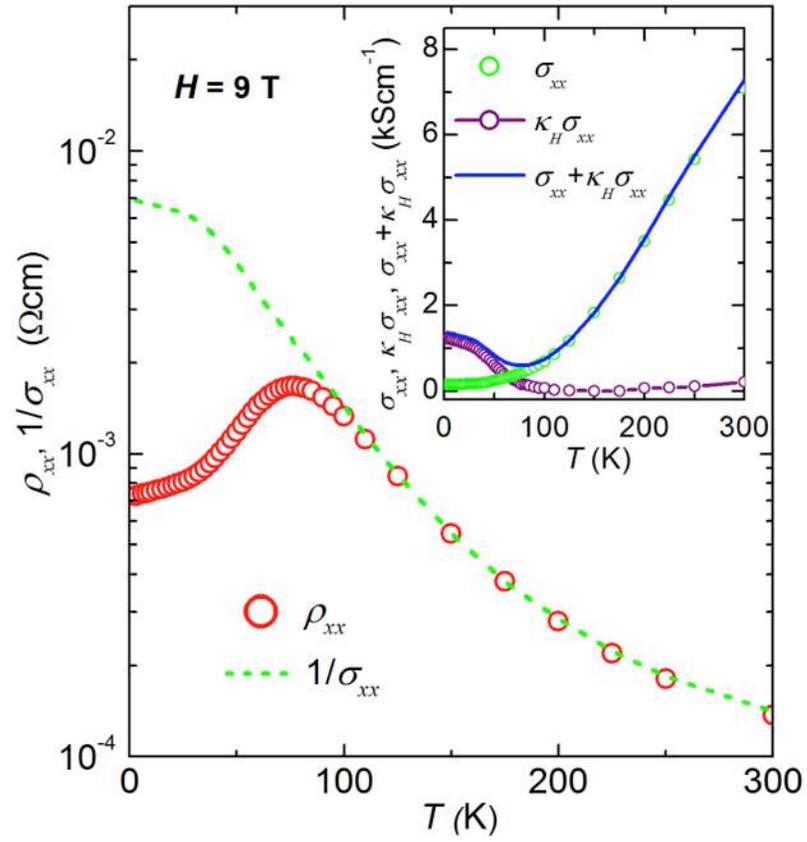

**FIGURE 4**

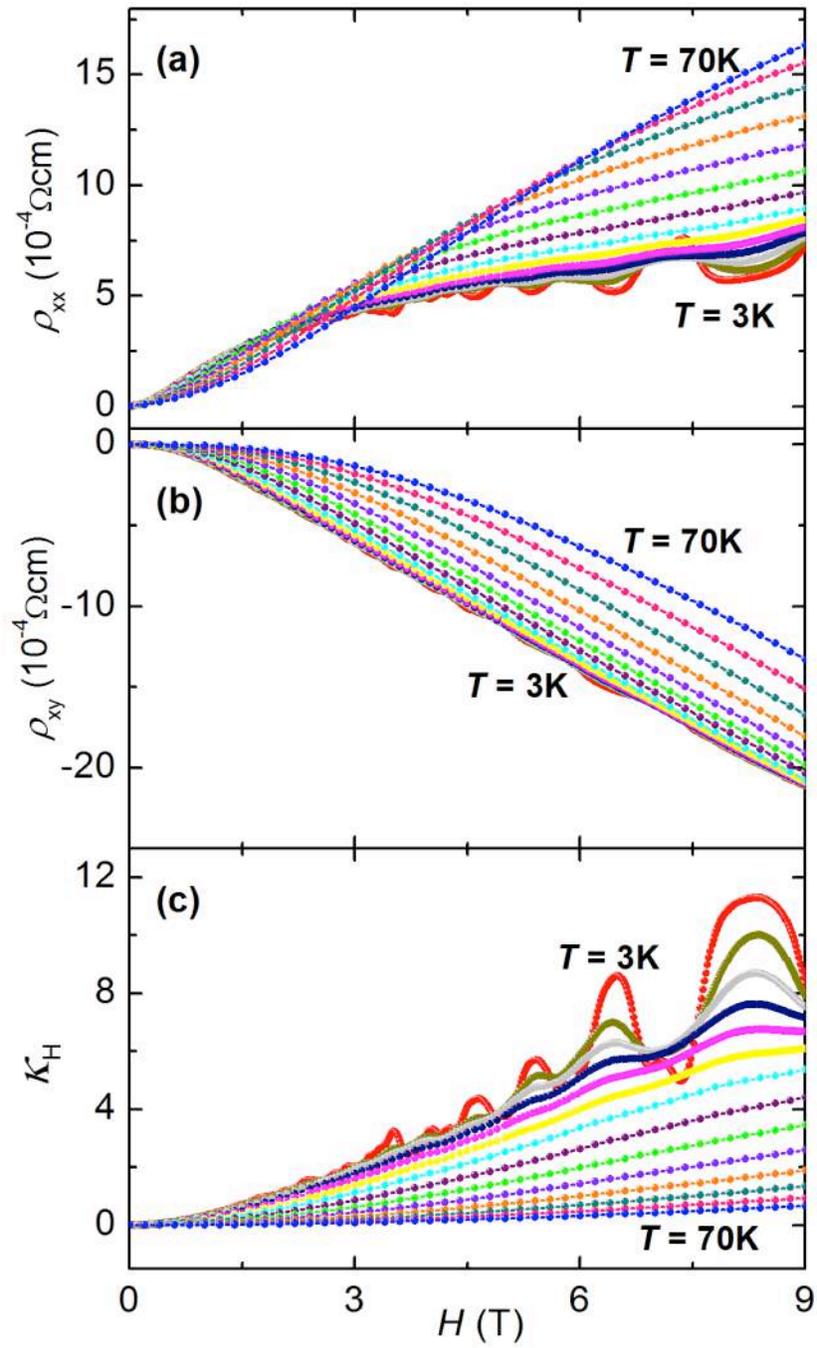

**FIGURE 5**

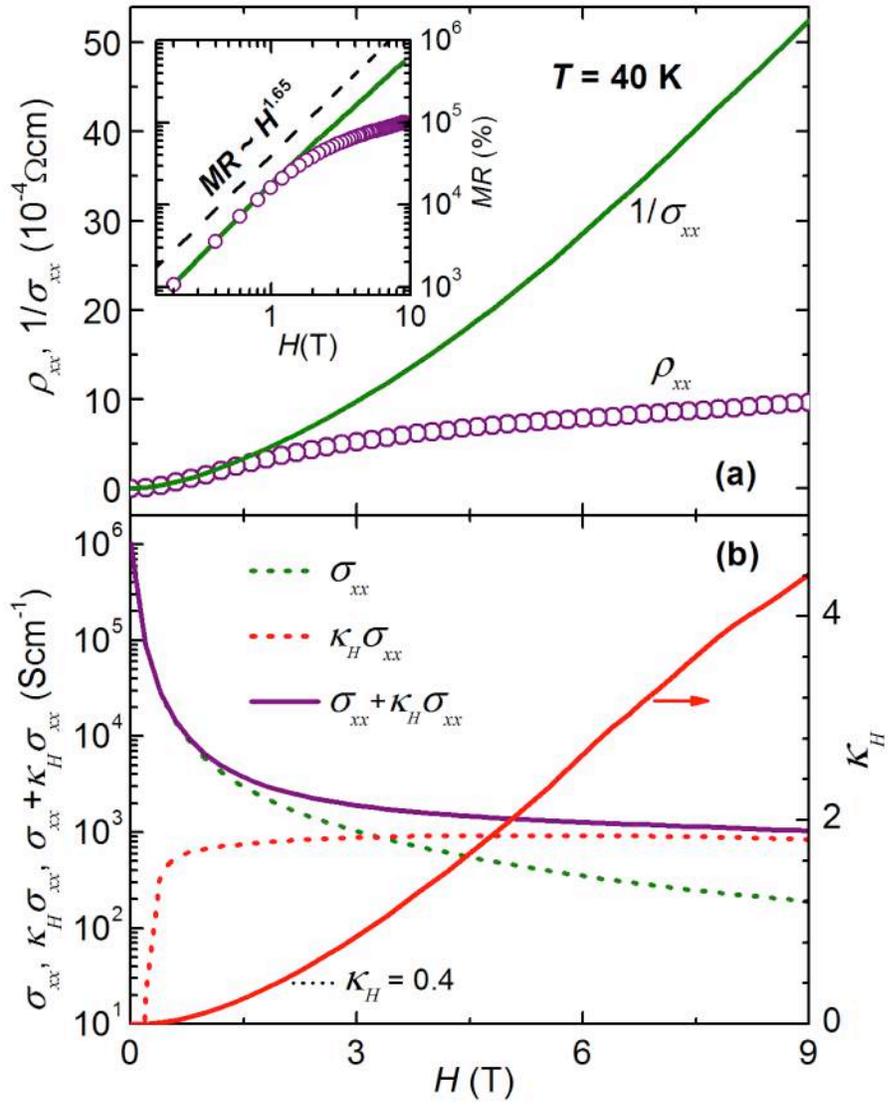